# Measurement-based quantum convolutional neural network for deep learning

Yifan Sun and Xiangdong Zhang*

Key Laboratory of advanced optoelectronic quantum architecture and measurements of Ministry of Education,

Beijing Key Laboratory of Nanophotonics & Ultrafine Optoelectronic Systems, School of Physics, Beijing

Institute of Technology, 100081, Beijing, China

*Author to whom correspondence should be addressed. E-mail: zhangxd@bit.edu.cn

## Abstract

Recently, quantum convolutional neural networks (QCNNs) are proposed, harnessing the power of quantum computing for faster training compared to the classical counterparts. However, this framework for deep learning also relies on multiple processing layers to capture the representation of data, which necessitates precise dynamical control. Given the current stage of quantum computing, achieving this level of control at a large scale remains challenging. Here, we propose an alternate approach to implementing QCNNs by utilizing cluster states. The training process of the method involves tuning the projection basis of each qubit in cluster states, rather than adjusting the parameters of layers of operators in deep quantum circuits. Hence, the whole system is easier to stabilize by avoiding the complex controls. Leveraging techniques in measurement-based quantum computing, we present an exact cluster state solution to general QCNNs. Followingly, we provide numerical evidence that both quantum and classical data can be learned by measuring cluster states, and a faster convergence of the method is observed. The cluster states we consider in our learning examples are merely square-lattice cluster states, whose implementation at large scale have been reported recently. It indicates that our method has the potential for realizing the advance of quantum deep learning for practical uses.

## I. INTRODUCTION

As the core of machine-learning technology, deep-learning methods have shown great power in aiding many aspects of modern society, such as web searches, image and speech recognition, etc [1–3]. The methods are generally performed by introducing multiple layers of modules for differently leveled representations. Starting with the raw input, they employ simple but non-linear structures that transform the representation at one level into a representation at a higher, slightly more abstract level. By using enough layers of such transformations, very complex functions of high-dimensional data can be learned. Undoubtably, the most intriguing method of deep learning by far is the convolutional neural network (CNN). The basic structure of a CNN contains an input layer, an output layer and multiple hidden layers for representations of data. The hidden layers include convolutional layers, pooling layers, normalization layers, and a fully connected (FC) layer. The convolutional layer applies a convolutional operation to map the input to its feature at certain level. Pooling layer is used to reduce the dimensionality, by associating the output from many neurons to fewer neurons. The FC layer connects every neuron of one layer to that of another. Actually, such a combination of the different kinds of layers have been proven to be effective in representing many kinds of data, leading to wide applications in activity recognition [4,5], sentence classification [6], text recognition [7], face recognition [8], object detection and localization [7,9], image characterization [10], etc.

Despite the great achievements by CNNs and other deep learning techniques, the rapid growth of data brings new challenge. The computing resources for training deep learning models on large datasets has also grown largely with the data, and is approaching to its saturation at current stage. This leads to the development of other computing formulism beyond the current one, and quantum deep learning [11–16] is one of the promising candidates [17–24]. Excitingly in this direction, several strategies have shown potential in fulfilling the needs for fast learning on given datasets. At present, a series of quantum learning algorithms are proposed, including the quantum generative network [12], quantum deep transfer learning [25], etc [20,26]. Very recently, the quantum convolutional neural network (QCNN) is theoretically constructed and investigated [27–30]. Such a network has displayed its unique property for identifying the symmetry of quantum states. More importantly, it has been numerically shown that the convergence of a QCNN model is faster than the CNN model in the task of classifying the classical data [28], and effectively reduce the barren plateau which causes troubles in quite a number of quantum neural networks [31], etc [32]. The basic scheme of QCNN is shown in Fig. 1(a). Like a CNN, it also contains convolutional layers, pooling layers, and an FC layer, except that all of the components are defined by quantum computing theory. The convolution layer in QCNN employs a series of two-qubit unitary operators $U_{ij}$ on adjacent qubits, with integers $i$ and $j$, as shown in the blue area in Fig. 1(a). The pooling layer is constructed by a series of control rotation gates. Each of the gates performs unitary rotations denoted by $V_{ij}$ on the target qubits, and the control qubits are then removed, as shown by the yellow area in Fig. 1(a). After layers of convolution and pooling, the size of state encoding the data is extensively suppressed in comparison with that of the input state. Then, an FC layer $F$ is applied to the remaining qubits, as shown by the green area in Fig. 1(a). It is a unitary gate with no specific requirements and can be adjusted according to the given task. Finally, the outcome of the circuit is obtained by measuring the output qubits. In particular learning tasks, the above QCNN processes the information encoded by qubits, whose parameters minimize the loss function after training. It can be seen that the learning enabled by QCNN shown in Fig. 1(a) is also based on layered structures. However, layered structure would cause troubles. After being operated by layers of gates, the qubits of QCNN or other deep quantum circuits are required to maintain the coherence. Meanwhile, because the error of the gates accumulates through the deep circuits, the single gate must be implemented with high enough fidelity in order to keep the final outcome reliable. These two requirements are still challenging for current quantum techniques [33]. Hence, it is general hard to realize the quantum machine learning strategy like QCNN at large scale for practical use.

In this work, we propose a different way to perform deep learning on the basis of quantum computing, by only measuring a quantum cluster state rather than applying layered structures. A cluster state is a special type of quantum entangled state that can performs universal quantum computing by local measurements. Such a computing strategy is termed by measurement-based quantum computing (MBQC) [34–46]. The particular techniques we used for our scheme are from MBQC, so we call it the measurement-based quantum convolutional neural network (MBQCNN). The key ingredient of our scheme also includes only two parts: preparing a proper cluster state, and performing the local measurements on the state. The final outputs of the learning process are given by the measurements results of the cluster states. In the numerical study, we focus on applying square lattice cluster states for the learning tasks, which do not show a one-to-one correspondence with the QCNN circuit, and we find their performance are quite satisfactory. Like the QCNN, they can display a faster convergence of the loss function over the classical CNN. Moreover, they can even provide an improvement in learning accuracy over the QCNN for certain cases. Recently, several investigations have demonstrated that very large cluster state can be prepared experimentally [47–49]. Therefore, our proposal has the potential to be realized at large scale, which may be applied to various practical datasets.

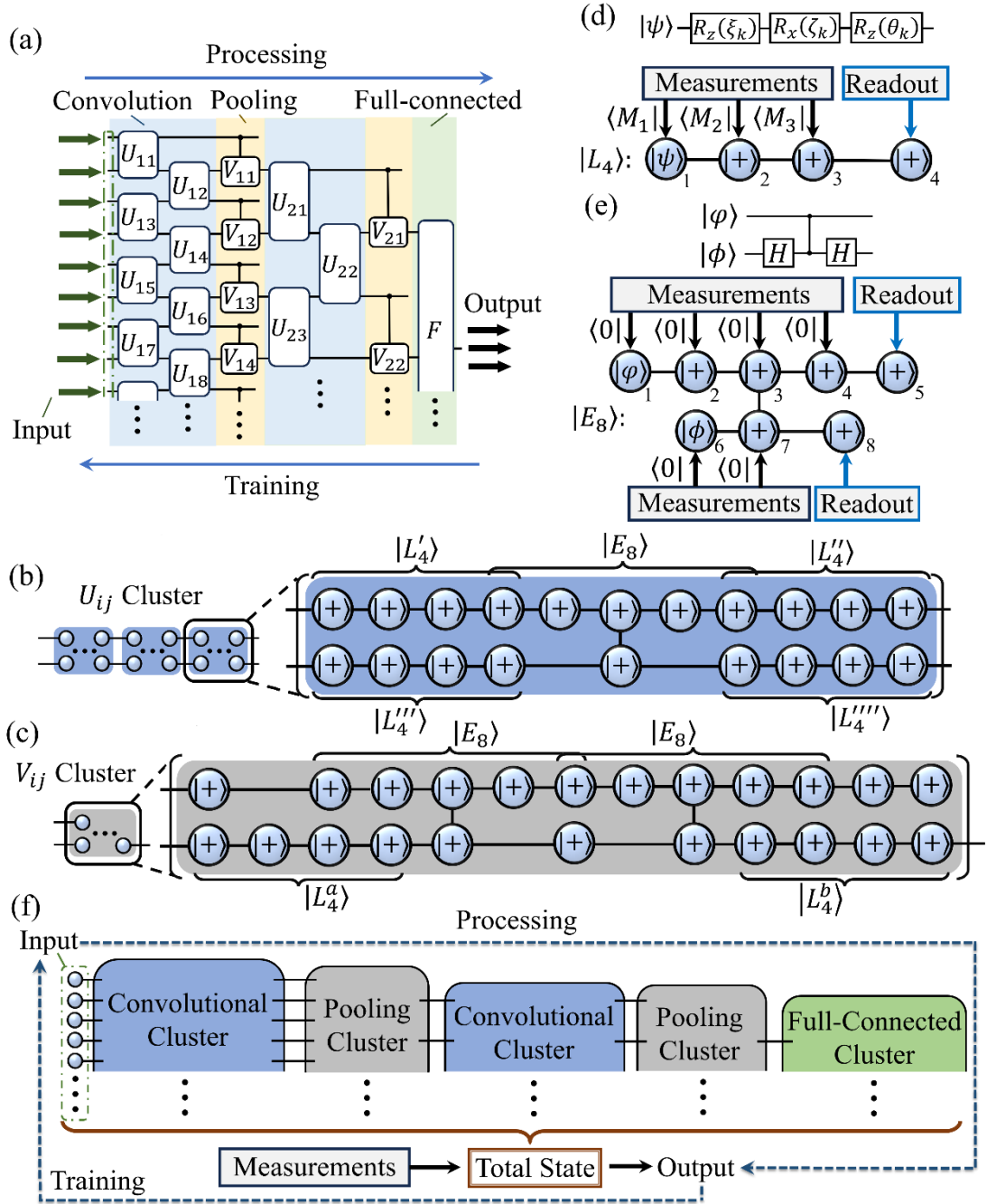

FIG. 1. Schematics of QCNN and the exact measurement-based scheme for performing its function. (a) General scheme of QCNN. The convolutional layer and pooling layer are marked by blue and yellow respectively. The FC layer is marked by green. The procedure of processing the data is indicated by the right-to-left arrow in the top. The procedure of training is indicated by the left-to-right arrow in the bottom. (b) The $U_{ij}$-cluster state. It is composed of three structurally repetitive blocks, each of which is built by four $|L_4\rangle$s (denoted by $|L_4\rangle$, $|L_4'\rangle$, $|L_4''\rangle$, and $|L_4'''\rangle$) and one $|E_8\rangle$. (c) The $V_{ij}$-cluster state. It is composed of two $|L_4\rangle$s (denoted by $|L_4^a\rangle$ and $|L_4^b\rangle$) and two $|E_8\rangle$s ($|E_8\rangle$ and $|E_8'\rangle$), together with three extra qubits. (d) The details of $|L_4\rangle$, which performs single qubit rotations (shown upward) after being measured. The first qubit in $|L_4\rangle$ serves as the input state, denoted by $|\psi\rangle$. By measuring the first to third qubits using $\langle M_1|$, $\langle M_2|$, and $\langle M_3|$, the fourth qubit is transformed to the rotation output. (e) The details of $|E_8\rangle$, which performs

CNOT operator (shown upward) after being measured. The first and sixth qubits serve as the input states, denoted by $|\varphi\rangle$ and $|\phi\rangle$. By measuring all the qubits except the fifth and the eighth ones using $\langle 0|$, the fifth and the eighth qubits are transformed to the CNOT output. (f) The total cluster state for performing QCNN. It is composed of sub-clusters, including convolutional clusters, pooling clusters, and a FC cluster. The sub-clusters are built by connecting $U_{ij}$-clusters or $V_{ij}$-clusters like the $U_{ij}$ gates or $V_{ij}$ gates of QCNN circuit in (a). The FC cluster is also a small-scale cluster state for performing unitary gate, which can also be given by $|L_4\rangle$ and $|E_8\rangle$. The input is set by connecting the qubits that encodes the data to a convolutional cluster. The processing of the data is indicated by the arrow in the top. The training of the model is indicated by the arrow in the bottom.

## II. THE MEASUREMENT-BASED SCHEME FOR QCNN

In contrast to the QCNN based on layers, we firstly construct a cluster state that has a strict correspondence with the QCNN circuit, by properly connecting the sub-cluster states like convolutional clusters, pooling clusters, and FC clusters. Each sub-cluster state performs the same function with the components of QCNN circuit after being measured. For example, measuring convolutional cluster implements the function of convolutional layer of QCNN circuit, and the same for others. The sub-clusters are further built by $U_{ij}$-clusters and $V_{ij}$-clusters, as shown in Fig. 1(b) and 1(c). They are set to perform the function of unitary operators and control rotation operators in Fig. 1(a).

Particularly, a $U_{ij}$-cluster performs the function of $U_{ij}$ gate in the convolutional layer of QCNN circuit. As shown in Fig. 1(b), the graph of $U_{ij}$-cluster is composed of three repetitive sub-graphs. Each subgraph is made by connecting the graphs of five special cluster states $|L_4'\rangle$, $|L_4''\rangle$, $|L_4'''\rangle$, $|L_4''''\rangle$, and $|E_8\rangle$. The four states $|L_4'\rangle$, $|L_4''\rangle$, $|L_4'''\rangle$, and $|L_4''''\rangle$ belong to the same type of cluster states represented by $|L_4\rangle$ [50]. As shown in Fig. 1(d), $|L_4\rangle$ is composed of four qubits, which are connected one-by-one into a line, and performs the single qubit rotation. To achieve so, the first qubit is set to be the input qubit state $|\psi\rangle$, and the rest qubits are set to be $|+\rangle = (|0\rangle + |1\rangle)/\sqrt{2}$, as marked in Fig. 1(d). The connections between the qubits represent the control-Z operator. Then, $|L_4\rangle$, is expressed by

$$|L_4\rangle = U_{CZ}^{1\to 2} U_{CZ}^{2\to 3} U_{CZ}^{3\to 4}(|\psi\rangle_1 \otimes |+\rangle_2 \otimes |+\rangle_3 \otimes |+\rangle_4), \tag{1}$$

where $U_{CZ}^{p\to q}$ are control-Z operator with the $p$th qubit being the control qubit and the $q$th qubit being the target qubit. If the measurement basis of the first three qubits is set to be $\langle M_1| = \langle 0|R_z(\theta_k)$, $\langle M_2| = \langle 0|R_z(\zeta_k)$, and $\langle M_3| = \langle 0|R_z(\xi_k)$, one has

$$(\langle M_1| \otimes \langle M_2| \otimes \langle M_3| \otimes I)|L_4\rangle = U_H R_z(\xi_k) R_x(\zeta_k) R_z(\theta_k)|\psi\rangle_4, \tag{2}$$

where $R_x$ and $R_z$ represent Pauli-X and -Z rotations. $|\psi\rangle_4$ in the right side of Eq. (2) indicates that the readout is obtained by the fourth qubit as shown by the rightmost one in Fig. 1(d). The redundant Hadamard gate operator $U_H$ can be absorbed into the rotation by a constant angle. According to Eq. (2), the rotation angles can be tuned by setting the measurement direction of the basis. The four states $|L_4'\rangle$, $|L_4''\rangle$, $|L_4'''\rangle$, and $|L_4''''\rangle$ are of the same form with $|L_4\rangle$, but with different inputs and independent measurement basis. Therefore, we apply four different notations. The state $|E_8\rangle$ is utilized to perform the function of a CNOT gate, and the details of the scheme is given by Fig. 1(e). To achieve the CNOT output, the first and the sixth qubits of $|E_8\rangle$ are set to be the input two-qubit state $|\varphi\rangle \otimes |\phi\rangle$, and the rest qubits are also set to be $|+\rangle$, as marked in Fig. 1(e). Thus, $|E_8\rangle$ is expressed as

$$|E_8\rangle = U_{CZ}^{1\to2} U_{CZ}^{2\to3} U_{CZ}^{3\to4} U_{CZ}^{4\to5} U_{CZ}^{3\to7} U_{CZ}^{6\to7} U_{CZ}^{7\to8}(|\varphi\rangle_1 \otimes |+\rangle_{2,5} \otimes |\phi\rangle_6 \otimes |+\rangle_{7,8}), \quad (3)$$

where $|+\rangle_{2,5} = |+\rangle_2 \otimes |+\rangle_3 \otimes |+\rangle_4 \otimes |+\rangle_5$, and $|+\rangle_{7,8} = |+\rangle_7 \otimes |+\rangle_8$. If the measurement basis of the qubits is set to be $\langle 0|$ except the fifth and the eight qubits, one has

$$(\langle 0|_{1,4} \otimes I \otimes \langle 0|_{6,7} \otimes I)|E_8\rangle = U_{CNOT}(|\varphi\rangle_5 \otimes |\phi\rangle_8), \quad (4)$$

where $\langle 0|_{1,4} = \langle 0|_1 \otimes \langle 0|_2 \otimes \langle 0|_3 \otimes \langle 0|_4$, and $\langle 0|_{6,7} = \langle 0|_6 \otimes \langle 0|_7$. Like Eq. (2), $|\varphi\rangle_5$ and $|\phi\rangle_8$ in the right side of Eq. (4) indicates that the readout is obtained by the fifth qubit and the eighth qubit as shown by Fig. 1(e). Under the above setup, the four cluster states $|L_4'\rangle$, $|L_4''\rangle$, $|L_4'''\rangle$, and $|L_4''''\rangle$ are connected to the first, fifth, sixth, and eighth qubit of $|E_8\rangle$ respectively, by control-Z operators, and the graph of the connected state is given by the panel in square bracket of Fig. 1(b). As shown by Fig. 1(b), three cluster states represented by the panels are joint together to give the $U_{ij}$-cluster. The measurement strategy of the state is set according to the function of each $|L_4\rangle$ and $|E_8\rangle$. In particular, the part in form of $|L_4\rangle$ is measured by $\langle M_1| \otimes \langle M_2| \otimes \langle M_3|$ with all angles being freely tuned. The part in form of $|E_8\rangle$ is measured by $\langle 0|$. Thus, the total measurement basis is the tensor product of $\langle M_1| \otimes \langle M_2| \otimes \langle M_3|$ and $\langle 0|$. By setting the leftmost two qubits of a $U_{ij}$-cluster (i.e., the first qubits of two $|L_4\rangle$s) to be the input qubits and applying the above measurement basis, the rightmost two qubits (i.e., the fourth qubits of two $|L_4\rangle$s) are the states corresponding to the two-qubit unitary transformation of the input qubits. The above construction of a $U_{ij}$-cluster is guaranteed by decomposition of a general two qubit unitary, which is thoroughly given in Supplementary Information S1.

A $V_{ij}$-cluster performs the function of the control-$V_{ij}$ gate in the pooling layer of QCNN circuit, and can be given in the similar manner. As shown in Fig. 1(d), the graph of $V_{ij}$-cluster is built by connecting two $|L_4\rangle$s and two $|E_8\rangle$s [50], and the connection is also enabled by a control-Z operator. The two $|L_4\rangle$s are independently measured, so they were denoted by different superscripts $a$ and $b$. Followingly, a $V_{ij}$-cluster can be obtained by connecting the $|L_4^a\rangle$ and $|L_4^b\rangle$ to the two qubits of $|E_8\rangle$s individually. One connected qubit is the input qubit of the first $|E_8\rangle$, and the other one is the output qubit of the second $|E_8\rangle$. There are three extra qubits. One of them is connected to the input of the first $|E_8\rangle$, and the other two are connected to the output of the second $|E_8\rangle$. They are applied for performing the function of $U_H$ and $R_z$. The measurement basis for a $V_{ij}$-cluster has some constrains. If the basis for $|L_4^a\rangle$ is $[\langle 0|R_z(\alpha)] \otimes [\langle 0|R_z(\beta)] \otimes [\langle 0|R_z(\gamma/4)]$, the basis for $|L_4^b\rangle$ is required to be $[\langle 0|R_z(\gamma/4)] \otimes [\langle 0|R_z(-\beta)] \otimes [\langle 0|R_z(-\alpha)]$, and the basis for the shared qubit of $|E_8\rangle$ is required to be $[\langle 0|R_z(-\gamma/2)]$. Besides, the basis for the single qubit connected to the first $|E_8\rangle$ is $\langle 0|$, and the basis for the two-qubit line state connected to the second $|E_8\rangle$ is $[\langle 0|R_z(\alpha)] \otimes \langle 0|$. By setting the leftmost two qubits of a $V_{ij}$-cluster to be the input qubits and applying the above measurement basis, the rightmost two qubits are the states corresponding to the control-rotation output of the input state. The above construction of a $V_{ij}$-cluster is guaranteed by decomposition of a general two qubit control-rotation gate. A detailed description is given in Supplementary Information S2.

By combining $U_{ij}$-clusters and $V_{ij}$-clusters, convolutional clusters and pooling clusters can be further given. Particularly, convolutional clusters are constructed by connecting $U_{ij}$-clusters in a zig-zag way, like the $U_{ij}$ gates of the QCNN circuit in Fig. 1(a). In one convolutional cluster, two output qubits of different $U_{ij}$-clusters are connected to the input qubits of a single $U_{ij}$-cluster, so that the entanglements of the cluster state can be relatively complex and effective for machine learning. Pooling clusters can be constructed by parallelly arranging the $V_{ij}$-clusters, like the control-$V_{ij}$ gates of the QCNN circuit in Fig. 1(a). They are applied to compress the data flow as the pooling layer in QCNN, linking a large convolutional cluster to a smaller one. By connecting those cluster states, as shown in Fig. 1(f), the total cluster state for performing the function of QCNN can be obtained. The measurement basis for the total state is the tensor product of the local projectors described by Eq. (2), Eq. (4), and the relevant requirements mentioned above. Although the connections inside the total cluster state are topologically similar to the connections among gates in a QCNN circuit, the two learning strategies are physically different. In the learning strategy based on QCNN circuits, the data encoded by qubits evolves under the quantum control, and the lines connecting the gates represent evolutions involving only global phases. In the learning strategy based on MBQCC, as we proposed, the data encoded by qubits are connected to the prepared total state, and the connection lines inside the total states represent control-$Z$ operators between the qubits. Then, the output can be obtained by merely measuring the total state. The training of MBQCNN is done by tuning the parameters of the measurement basis. Notice that, in a strict sense, the constrains for $V_{ij}$-clusters mentioned above is required to keep during the training. However, in the numerical results below, we show that they are unnecessary for practical cases, when only the effectiveness of the model is important. Like the training of QCNN circuit, an MSE loss function can be defined in the task of learning data set $\{x_n, L_n: n = 1, \dots, N\}$, where $x_n$ is the $n$th data vector and $L_n$ is the label of it. Suppose that the total cluster state for MBQCNN is denoted by $|T\rangle$, and the measurement basis is denoted by $\langle M_{total}|$. The MSE loss can be expressed by

$$\frac{1}{2N}\sum_{n=1}^{N}\left[L_n - \left|\langle M_{total}|\widetilde{U}_{CZ}(|\pmb{x}_n\rangle \otimes |T\rangle)\right|^2\right]^2, \quad (5)$$

where $\widetilde{U}_{CZ}$ is the operator that connects input state $|\pmb{x}_n\rangle$ to $|T\rangle$. Similar to other training processes, the values of parameters of the measurement basis finally minimize Eq. (5), implying the terminal of the training. Each parameter can be updated by gradient descent, back-propagation, or other common methods for model training. The scheme of MBQCNN indicates that deep learning enabled by QCNN can also be performed by a strategy without deep networks. Furthermore, the long-time dynamical control required by the QCNN circuit may not be a technique issue for performing quantum deep learning at large scale. Such dynamical controls can be effectively implemented by just measuring cluster states.

### III. LEARNING EXAMPLES BY MBQCNN

In this section, we give two examples to show the performance of MBQCNN. The task of the first example is to identify the quantum phase of Haldane ground states [51]. The Haldane Hamiltonian on the spin-1/2 chain with open boundary conditions is defined by

$$H = -J\sum_{i=1}^{N-2} Z_i X_{i+1} Z_{i+2} - h_1 \sum_{i=1}^{N} X_i - h_2 \sum_{i=1}^{N-1} X_i Z_{i+1}, \quad (6)$$

where $X_i$ and $Z_i$ are Pauli operators for the $i$th spin. $h_1$, $h_2$, and $J$ are the parameters of the Hamiltonian. The ground states of the Hamiltonian can show special symmetries, which can be justified by the non-zero string order parameters (SOPs) such as $S_{mn} = Z_m X_{m+1} ... X_{n-1} Z_n$ [27,51]. In Ref. [18], the task has been tackled by the QCNN circuit model. Their results show that the ground states processed by a well-trained QCNN circuit are more convenient for quantum phase identification, because the outputs of the circuit are low dimensional states whose correlation are relatively easy to measure. Hence, an advance in the number of particles that is required to measure is observed. Here, we apply the MBQCNN strategy for tackling the task. The data set we consider is the same with Ref. [18], each sample of which is composed of a ground state $|g_i\rangle$ labeled by its SOP expectation. The cluster state we consider for the task is a little different from the state in Fig. 1(f). In our proposal for constructing the total cluster state in Fig. 1(f), there is a clear relation between each block of the cluster state and the layer in a QCNN circuit. However, the general unitary gates can be implemented by multiple types of cluster states, such as the square-lattice states which is experimentally implementable at large scale [47–49]. Hence, the function of the convolution clusters can be generally implemented with square-lattice states, and the function of the pooling clusters can be generally implemented by measuring partial qubits of square-lattice states. Then, in specific cases, the complex connections in the total cluster state can be simplified into a regular type such as the square-lattice. The measurement strategy also needs to reform. In the measurement setup for the total state in Fig. 1(f), some qubits are measured by fixed basis, such as those composing the substate $|E_8\rangle$. In the measurement setup for the square-lattice state, each qubit is required to be measured by a parameterized basis, denoted by $\langle 0|(R_y R_z)$ with Pauli-$Y$ and Pauli-$Z$ rotation $R_y$ and $R_z$. This can be considered as a trade-off between the scheme strictly related to the QCNN circuit and the scheme only giving the equivalent output.

The specific case we consider is when $N = 3$. In such a case, the input data sample $|g_i\rangle$ is encoded by three extra qubits, and the cluster state for the task is set to be a 2-by-5 square lattice cluster state. The input process of the data sample is done by connecting the three qubit states to the middle three qubits of the square lattice boundary containing five qubits. This gives a 13-qubit state, denoted by $|S_{13}(i)\rangle$. The reason for choosing such a state is that it basically has a 3-qubit "branch" serving as the input port, and the rest qubits of it are sufficient to perform 3-qubit unitary operators on the input. The QCNN circuit for the task of the same size contains only two unitary gates and two control-rotation gates. The details of the circuit and the square lattice state is graphically given in Supplementary Information S3. The output of the scheme is given by the measurements,

$$\Omega(\boldsymbol{\alpha}, \boldsymbol{\beta}) = \left| [\langle 0|R_y(\alpha_1)R_z(\beta_1) \otimes \langle 0|R_y(\alpha_2)R_z(\beta_2) \otimes \cdots \otimes \langle 0|R_y(\alpha_{13})R_z(\beta_{13})]|S_{13}\rangle \right|^2, \quad (7)$$

where $\boldsymbol{\alpha} = (\alpha_1, \alpha_2, \ldots, \alpha_{13})$ and $\boldsymbol{\beta} = (\beta_1, \beta_2, \ldots, \beta_{13})$ are real parameters. All the parameters are updated by the gradient of the MSE loss function, as defined by Eq. (5). In the numerical study, we consider three different-sized training datasets. Each dataset is sampled from the ground states of Eq. (7) under different $h_1$, $h_2$, and $J$s. Particularly, $h_1/J$ is set to be from 0 to 2, and $h_2/J$ is set to be from -2 to 2. The number of samples are $6 \times 6$, $9 \times 9$, and $12 \times 12$ respectively, under equal-spaced values of $h_1/J$ and $h_2/J$ in the above ranges. The testing datasets are obtained by similar procedure, but the samples of them are completely different from those in the training datasets. The number of samples in the testing datasets are around ten percent of the corresponding training datasets. The convergence curve of the training process is shown by Fig. 2(a). The upper three curves are the accuracies obtained by testing sets, and lower three curves are the loss function values obtained by the corresponding training sets. The curves of different-sized datasets are marked by different colors. From the results in Fig. 2(a), it can be concluded that loss/accuracy values in three cases all converges well. We further check the phase boundaries obtained by the output of $|S_{13}\rangle$, and the results are given by the blue curves in Fig. 2(b). The standard 3-qubit results are given by the black lines. The boundaries of phase are identified by the second order derivatives of the output or SOPs. The two curves match well, indicating that the 2-by-5 square lattice cluster state can be as effective as the QCNN circuit on the task of learning the characters of the Haldane ground states.

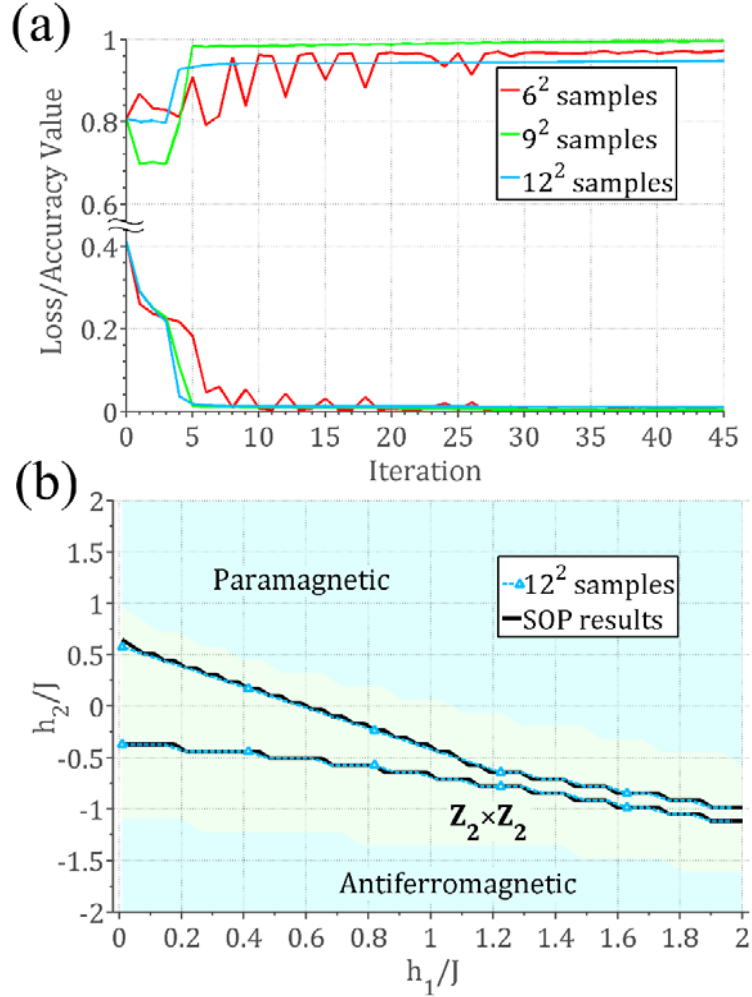

FIG. 2. The performance of MBQCNN when learning the characters of the Haldane ground states. (a) The convergence of the scheme. The results obtained by training datasets of $6^2$, $9^2$, and $12^2$ samples are colored by red, green, and blue, respectively. The upper three curves represent the accuracy of testing sets, which are obtained by picking out the ground states under different parameters from those in training datasets. The lower three curves represent the loss function value of training datasets. (b) The phase boundaries obtained by the MBQCNN, shown by the blue dashed curves with triangle markers. The black lines are the boundaries obtained by SOP. The two curves match well. The background marks the area of different phases, which is obtained by 12-site Haldane Hamiltonian.

The task of the second example is to classify the iris dataset. Such a dataset is composed of the 150 4-by-1 vectors [25,52], describing the petal and sepal information of three kinds of iris flowers. The illustration of the samples in iris dataset is given by Fig. 3(a), where each point in the figure represents a data sample. The

3D coordinates of a point represent the first three components of a data sample. The color of a point indicates the value of the fourth component, and specified by the color bar. From Fig. 3(a), one can see that there are no clear class boundaries for three kinds of iris flowers. So, target of the task is to classify each kind of flower by learning the dataset, which is a typical classification task.

Like the first example, we also employ a square lattice cluster state for tackling the task. The state is composed of a 2-by-4 square lattice with two extra qubits. Each extra qubit is connected to the one of the qubits in the cluster state boundaries involving two qubits. Besides, the diagonals of the 2-by-2 square in the other side is also connected for increasing the learning ability of the scheme. The reason for choosing such a state is similar. Firstly, it has a 4-qubit "branch" serving as the input port. Secondly, the rest of the cluster state contains the structure that connects a 4-qubit branch to a 2-qubit branch, which is also sufficient to perform special 4-qubit unitary operators on the input. A graphic illustration of the state is given in Supplementary Information S4. The input for the cluster state in this case is set to be a four-qubit state. In order to encode the 4-by-1 vectors of iris dataset, they are pre-processed into 16-by-1 vectors. The first four components of the new vectors are the same components of the original data samples, and the rest ones are set to be zeros. Because there are 50 samples for each kind of iris flowers, the training set is composed of picking out 40 samples from each kind, and the rest samples are employed for testing set. The labels of the three kinds of iris flowers are valued by 0, 0.5, and 1. The measurement basis is of the same form as described by Eq. (7). In order to check the performance, we compare the classification results of the above MBQCNN scheme with those obtained by a QCNN circuit model and a classical CNN model. The details of the QCNN and the CNN model we use is given by Appendix A and B. For a fair comparison, the parameter numbers of the QCNN and the CNN model equal to that of the MBQCNN, which is 28. The convergence of the training process is shown by Fig. 3(b). The upper three curves represent the accuracy obtained by the testing set, and the lower three curves represent the loss function value obtained by the training set. The results obtained by CNN, QCNN, and MBQCNN are colored by red, green, and blue, respectively. The training of the MBQCNN is the same with that in the first example. The training of the CNN and the QCNN is also done by gradient descent. We also introduce fluctuations to the training process, and the curves are obtained by averaging over five times of training. The variance is also marked by the area of the same corresponding color around the curves. From these results, we can see that the MBQCNN scheme shows the same speed of convergence with the QCNN circuit, which are faster than that of the CNN. The reason under the speed-up can be seen more clearly from the average of gradient magnitudes, as shown in Fig. 4. The average of gradient is a useful tool for estimating the trainability of models [31,53]. It reflects the average "step length" of the loss function towards a specific value, so that the duration of training is roughly proportional to its reciprocals. Because the parameter number of the three models are all 28, one can compute the gradients of the loss functions, which are 28-dimensional vectors. By averaging the magnitudes of the gradients over possible choices of parameters, an overall evaluation of the magnitude can be obtained. In order to show that sufficient choices of parameters are considered, the number of possible parameter choices is increased gradually, and the averages shown in Fig. 4 finally converge. It can be seen that, for such a dataset, the MBQCNN presents the best average magnitude of gradient among the three models, which is close to that of QCNN. The CNN in this case does not present a significant gradient, leading to a relatively long training process.

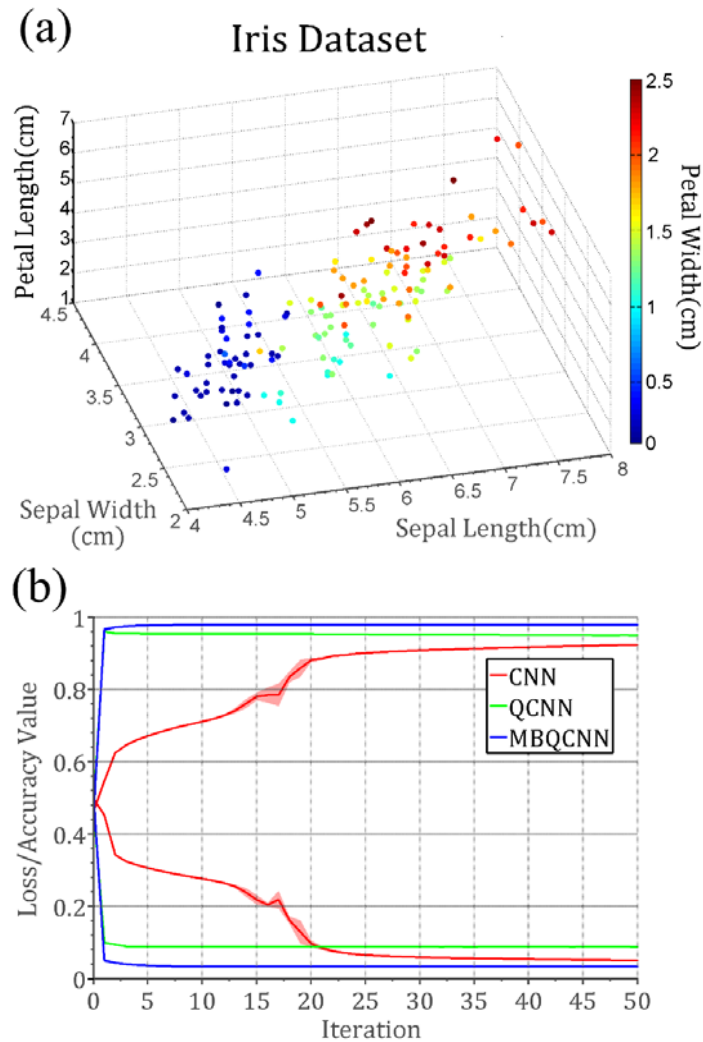

FIG. 3. The performance of MBQCNN scheme when classifying the iris dataset. (a) The illustration of the samples in the iris dataset. The 3-D coordinates of a point represent the first three components of a data sample. The color of a point indicates the value of the fourth component, and specified by the color bar. (b) The convergence of different schemes in the classification task. The results obtained by CNN, QCNN, and MBQCNN are colored by red, green, and blue, respectively. The upper three curves represent the accuracy of testing sets. The lower three curves represent the loss function value of training datasets. All the curves are obtained by averaging over five times of training, and the variance is marked by the areas near the curves which are colored accordingly.

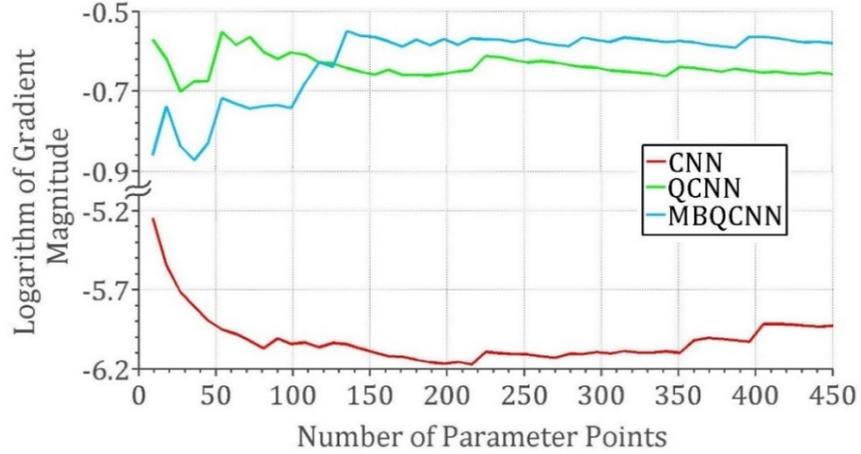

FIG. 4. The average gradient magnitude of the CNN, the QCNN, and the MBQCNN model we consider. The *y*-axis represents the logarithm of the average gradient magnitude with base 10. The *x*-axis represents the number of the choices of the parameters (or the number of parameter points for short, as denoted in the figure) that is taken for the average. All the parameters for the three models are randomly chosen from the interval $[0, 2\pi]$. The curves for the results of the CNN, the QCNN, and the MBQCNN are colored red, green, and blue, respectively.

The comparison between the two quantum schemes also shows that the final classification accuracy obtained by the MBQCNN scheme is higher than that obtained by the QCNN circuit. This is because the MBQCNN scheme here is not strictly equivalent to the QCNN circuit at the same scale, and it turns out to performs better. For general tasks, machine learning models are usually tuned structurally for better performance. Here in the MBQCNN scheme, the model can be tuned by adding or removing qubits and connections. It is different from re-arranging gates in the circuit model. Therefore, as verified by the numerical example, such a tuning strategy based on our new scheme for deep learning has a potential for achieving better outcomes.

## IV. EXPERIMENTAL REALIZATION

Recently, 2D square lattice states at large scale has been proven to be available via time-multiplexing optical parametric oscillators (OPOs) [47–49], and various types of measurements can also be realized. The results given by the two works actually provide ideal experiment platforms for our MBQCNN scheme. This can be seen by our numerical examples which is completely based on 2D square lattice cluster states. Although our scheme is based on discrete variable cluster state and the experimental platforms are for continue variable cluster state. In fact, the discrete variable cluster state as well as its operations can be implemented with continue variable cluster state, which has been widely discussed [35,38,40]. Therefore, the experimental scheme of our proposal is direct. We briefly introduce one of the strategies in Appendix C.

As for the measurement basis, it can also be implemented with the mentioned platform. In our proposal, the measurement basis is the tensor product of projector $\langle 0|R_z(\theta)$, or $\langle 0|\left(R_y(\phi)R_z(\theta)\right)$. Those projections belong to the local measurements of qubits or qumodes, and has been investigated. If one considers using the correspondence between the qubits or qumodes for implementing the MBQCNN scheme, as we explained in Appendix C, the measurement with the above form of projectors can be implemented by tuning the transmissive/refractive index of the interferometer in the homodyne detection [47].

## DISCUSSION AND CONCLUSION

Along with the explosive growth of data, how to enhance the computing capability has become a vital problem. In recent years, the potential advantageous strategies such as quantum deep learning has drawn much attention. However, learning the complex character of the dataset is traditionally considered to be accomplished by layered structures. This leads to troubles for current quantum techniques. The requirements on the long coherence time and high gate fidelity severely limits the depth of an available quantum circuit. Therefore, many theoretical quantum deep learning algorithms with a significant depth has not been applied for practical dataset.

In this work, we propose an alternative way for overcome the above problem, by realizing quantum deep learning via measuring cluster states without algorithmic depth. We focus on the recently proposed QCNN circuit, and provide a strict cluster state that can realize the function of every detail of the circuit. In the numerical investigation, we consider two examples. The first one is learning the phase of Haldane ground states, and second one is classifying the iris dataset. In the first example, we find that, although not having a one-to-one correspondence with the standard QCNN circuit, the square lattice can show the same performance on detecting the phase of given ground states. In the second example, surprisingly, we show that the square lattice state performs even better than the QCNN circuit, achieving a higher accuracy of classification. An advance in the convergence speed over the classical CNN method is also observed. Our work provides an opportunity of employing the recent advance in large cluster state preparation for implementing quantum deep learning algorithms. By mapping the quantum deep learning algorithm to a cluster state, the depth of the circuit is correspondingly transformed to the scale of the cluster state. Therefore, a large cluster state is sufficient for realizing the function of the quantum deep learning algorithm with a significant depth. The method would promote the implementation of quantum machine learning for real applications.

## ACKNOLEDGEMENT

This work is supported by the National Key R & D Program of China under No. 2022YFA1404900 and the National Natural Science Foundation of China under (No.12104041, No. 11904022, & No. 12474355).

## APPENDIX A: THE CNN MODEL APPLIED TO CLASSIFYING IRIS DATASET

As the reference, we consider a simple CNN model, which is composed of one convolutional layer and one pooling layer. The convolutional layer in CNN model is a classical one, composed of five kernels. Because the input data is a 16-by-1 vector, each kernel is set to be a 3-by-1 vector. The convolution of one input vector with five kernels generate five 14-by-1 vectors. Then, each element of the five 14-by-1 vectors is activated by the sigmoid function. Then, by using average pooling, the five 14-by-1 vectors are transformed to five numbers. Followingly, the five number are weighted by two different sets of coefficients with biases, giving two numbers. After being activated by a sigmoid function, the two numbers are weighted by one final set of coefficients with bias and sigmoid function, which gives the final output. In order to obtain a fair comparison with the MBQCNN, we set the second element of the fifth kernel to be zero, and the rest of the elements of the kernels, the weights, and biases are all trainable parameters. As such, the total parameter number of the model is also 28, which equals to that of the MBQCNN.

The training of the model is done by gradient descent. The loss function is set to mean square error (MSE). By using finite difference as the approximation, the gradient of MSE is computed numerically. The update of the parameters is realized by adding the product of gradient and learning rates. The learning rates change with the gradient. If the gradient of the current step is less than that of the last step, the learning rate is a random number in $[0.25, 0.75]$. Otherwise, the learning rate is a random number in $[0.95, 1.45]$. The introduction of the randomness is for checking the sensitivity of training process to the parameters. A graphic illustration of the model is shown in Supplementary Information S4.

## APPENDIX B: THE QCNN MODEL APPLIED TO CLASSIFYING IRIS DATASET

The details of the QCNN model we applied is given as follows. It is composed of one convolutional layer and one pooling layer as defined in Fig. 1(a). In QCNN, the convolutional layer is composed of two-qubit unitary gates. Here, because the input is a four-qubit state, which encodes a 16-by-1 vector, we adopt three two-qubit unitary gates in the convolutional layer. They act on the first-second, second-third, and third-fourth qubits respectively. Each one of them is of the form

$$[R_y(\tau_1) \otimes R_z(\tau_2)U_H]U_{CNOT}[R_y(\tau_3) \otimes U_H]U'_{CNOT}[R_z(\tau_4) \otimes R_y(\tau_5)]U_{CNOT}$$

$$[R_z(\tau_6) \otimes R_z(\tau_7)R_y(\tau_8))]. \tag{B1}$$

Such a form belongs to a subset of general two qubit unitary gate, and is sufficient for certain learning tasks [28]. The pooling layer here is composed of two control rotation gates. The first and third qubit are the control qubit, and the second and the fourth qubit are the target qubit. Each control rotation gate is of the form

$$[I \otimes R_z(\tau'_1)R_y(\tau'_2)]U_{CNOT}[I \otimes R_z(-\tau'_1)R_y(-\tau'_2)]U_{CNOT}. \tag{B2}$$

The final output is given by measuring the target qubits in basis state $|+\rangle \otimes |+\rangle$. Hence, the total parameter number of the model is also 28, which also equals to that of the MBQCNN.

The training of the model is the same with that of the CNN. The loss function is also set to be MSE. Particularly, the computation of the gradient, the update of the parameters, and the setup of the learning rate are the same with the description in Appendix A. A graphic illustration of the model is also shown in Supplementary Information S4.

## APPENDIX C: AN EXPLANATION OF CV CLUSTER STATE

In the main text, we mainly consider the qubit cluster state. Each node in the cluster state represents a qubit state $|+\rangle$, as shown in Fig. 1. The connection between each pair of nodes represent a CZ operator between the corresponding two qubits. The CV cluster state entangles qumodes [40], which are usually the zero-momentum eigenstate $|0\rangle_p$. The "CZ operator" entangling the $i$th and $j$th qumodes are expressed by $e^{iq_iq_j}$, where $q_i$ and $q_j$ are the position operator. The $X$ and $Z$ measurements for qubit cluster state computing corresponds to the measurements of momentum operator $\hat{p}$ and position operator $\hat{q}$ for CV cluster state computing. The operator $R_x$ and $R_z$ for qubit state corresponds to $\hat{X}(s) = e^{is\hat{p}}$ and $\hat{Z}(s) = e^{is\hat{q}}$ for CV state, respectively.

# Supplementary Information of Measurement-based quantum convolutional neural network for deep learning

Yifan Sun and Xiangdong Zhang*

*Key Laboratory of advanced optoelectronic quantum architecture and measurements of Ministry of Education, Beijing Key Laboratory of Nanophotonics & Ultrafine Optoelectronic Systems, School of Physics, Beijing Institute of Technology, 100081 Beijing, China.*
*Author to whom any correspondence should be addressed: zhangxd@bit.edu.cn

## S1. The deduction of the $U_{ij}$-cluster

The deduction of the $U_{ij}$-cluster is given by employing the decomposition of the unitary gate. It is known that a two-qubit unitary gate can be decomposed into single-qubit unitary gates and CNOT gates. Then, by using the clusters state that can perform the function of single-qubit unitary gates and CNOT gates (Fig. 1(c) and 1(d) in the main text), the $U_{ij}$-cluster can be constructed. Specifically, the decomposition of a two-qubit unitary gates $U_{ij}$ can be given by [1]

$$U_{ij} = (R_7 \otimes R_8) U_{CNOT} (R_5 \otimes R_6) U_{CNOT} (R_3 \otimes R_4) U_{CNOT} (R_1 \otimes R_2), \quad (S1)$$

where $R_1$ to $R_8$ are single-qubit rotations. The circuit plot of Eq. (S1) is given by Fig. S1. In the main text, we have pointed out that the single qubit rotation gate can be implemented by measuring the line-shape cluster state $|L_4\rangle$ shown in Fig. 1(d), and the CNOT gate can be implemented by measuring the T-shape cluster state $|E_8\rangle$ [2]. Therefore, the function of the circuit in Fig. S1, consisting of four single-qubit rotations and one CNOT, can be implemented by connecting the corresponding clusters and performing the required measurements on it. The connecting scheme is shown by Fig. S2.

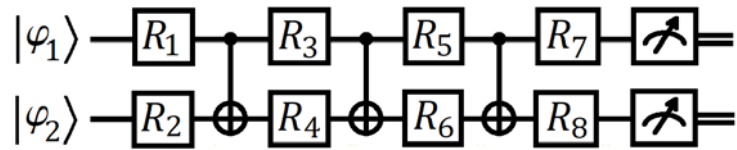

**Fig. S1** The circuit of a universal two qubit processor.

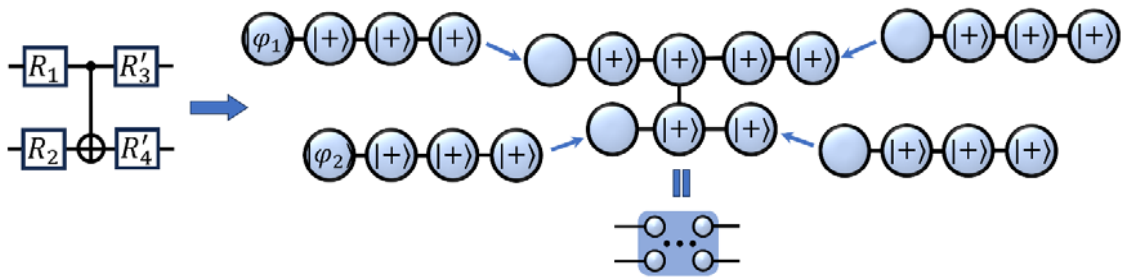

**Fig. S2** The method for obtaining the blocks of a $U_{ij}$-cluster.

Notice that the state obtained by Fig. S2 is the state shown in Fig. 1(b) in the main text. The four single qubit rotations implemented by measuring the four line-shape cluster states are denoted by $R_1$, $R_2$, $R'_3$, and $R'_4$ respectively. By applying three states shown in Fig. S2 and connecting them sequentially as shown in Fig. S3, the $U_{ij}$-cluster can be obtained. The corresponding single-qubit rotations implemented via measuring the three states are indicated through the gates of the circuits. Following Fig. S1, the rotation gates in the circuits of Fig. S3 satisfy that $R'_3 \cdot R''_3 = R_3$, $R'_4 \cdot R''_4 = R_4$, $R'_5 \cdot R''_5 = R_5$, and $R'_6 \cdot R''_6 = R_6$. These conditions imply that the measurement basis on the parts of the states for performing the single qubit rotations have to satisfy related constrains. Suppose that $R_3$ is expressed by a Z-X-Z decomposition, $R_z(\alpha)R_x(\beta)R_z(\gamma)$. The measurement basis for performing $R'_3$ and $R''_3$ are denoted by $\{\langle 0|R_z(\alpha'), \langle 0|R_z(\beta'), \langle 0|R_z(\gamma')\}$ and $\{\langle 0|R_z(\alpha''), \langle 0|R_z(\beta''), \langle 0|R_z(\gamma'')\}$ respectively. Then, the angle parameters satisfy

$$[R_z(\alpha')R_x(\beta')R_z(\gamma')] \cdot [R_z(\alpha'')R_x(\beta'')R_z(\gamma'')] = R_z(\alpha)R_x(\beta)R_z(\gamma), \qquad (S2)$$

which can be applied for setting the measurement basis. The rest of the constrains are similar to Eq. (S2).

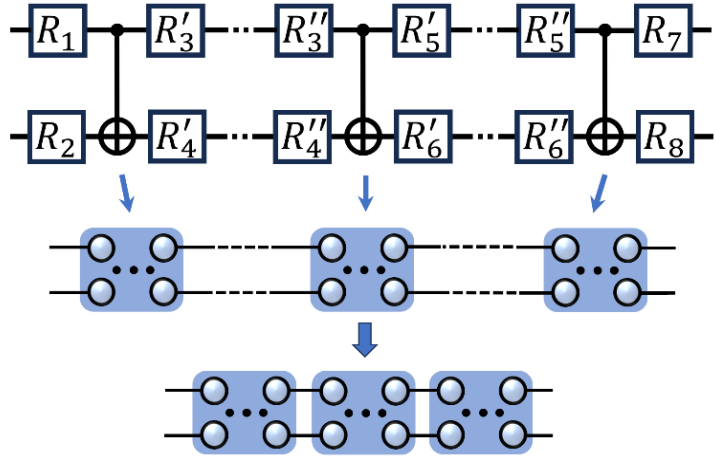

**Fig. S3** The method for obtaining a $U_{ij}$-cluster.

Additionally, we would like to point out that such a strategy is not optimal. Because a two-qubit unitary has only fifteen parameters, $R_3$ to $R_6$ need not to be arbitrary single qubit gate. According to the previous research [2], the optimal circuit for $R_3$ to $R_6$ are given by

$$R_3 = R_z\left(-2h_3 - \frac{\pi}{2}\right)U_H, \qquad R_4 = R_y\left(\frac{\pi}{2} + 2h_1\right)U_H,$$

$$R_5 = U_H, \qquad R_6 = U_H R_y\left(-2h_2 - \frac{\pi}{2}\right), \qquad (S3)$$

$U_H$ is the Hadamard operator, which can be expressed by $U_H = XR_y(\pi/2)$. $h_1$, $h_2$, and $h_3$ are three rotation angles. Following Eq. (S3), an optimized $U_{ij}$-cluster can be given by Fig. S4, in which the angles of the measurement basis are marked on the corresponding qubits. As pointed out in the main text, the local basis is of the form $\langle 0|R_z(x)$ ($x$ takes $a_{11}$, $a_{12}$, ..., $a_{43}$, $t_1$, $t_2$, and $t_3$). $t_1$, $t_2$, and $t_3$ satisfy that $t_1 = -2h_3 - \pi/2$, $t_2 = \pi/2 + 2h_1$, and $t_3 = -2h_2 - \pi/2$, respectively, and the rest of the angles $a_{11}$, $a_{12}$, ..., $a_{43}$ are set according to the rotations $R_1$, $R_2$, $R_7$, and $R_8$.

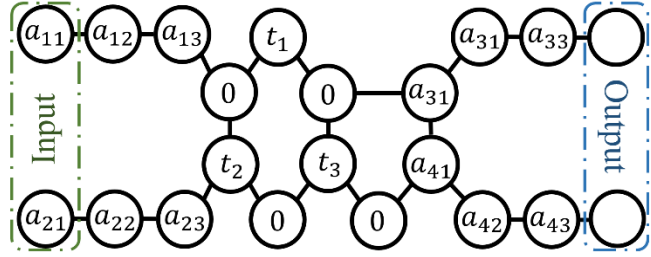

**Fig. S4** An optimized $U_{ij}$-cluster.

## S2. The deduction of the $V_{ij}$-cluster

The deduction of the $V_{ij}$-cluster is also based on a similar procedure as described in the above section. By employing the decomposition of a control rotation gate, the gate can also be given by a circuit composed of single-qubit rotation gates and CNOT gates. Then, the $V_{ij}$-cluster can be constructed by connecting the states in Fig. 1(c) and 1(d) in the main text according to the decomposition. The circuit of a general control rotation, with the control performed in the Pauli-X basis, can be given by Fig. S5 [1]. Particularly, the line-shape and the T-shape cluster states are connected according to the circuit in Fig. S5. Therefore, the circuit consisting of two single-qubit rotations (each of which is expressed by a $R_z$-$R_x$-$R_z$ decomposition), two $R_z$ rotations (each one serves as the gate $R_\alpha$, shifting the phase of $|1\rangle$ components by $\alpha$), two Hadamard gates, and two CNOTs can be given by connecting the corresponding clusters, as shown by Fig. S6. In the main text, the two line-shape states are denoted by $|L_4^a\rangle$ and $|L_4^b\rangle$, and T-shape states are denoted by $|E_8\rangle$s. Following the order number of the qubits in Fig. 1(d) and 1(c) of the main text, the connecting scheme can be described by a four-step procedure. First, connect the fourth qubit of $|L_4^a\rangle$ to the sixth qubit of a $|E_8\rangle$. Second, connect the fifth and eighth qubit of the $|E_8\rangle$ to the first and fifth qubit of another $|E_8\rangle$ respectively. Third, connect the eighth qubit of the last $|E_8\rangle$ to the first qubit of $|L_4^b\rangle$. Fourth, connect a single qubit to the first qubit of the first $|E_8\rangle$ and connect a two-qubit line state to the fifth qubit of the second $|E_8\rangle$. Then, the $V_{ij}$-cluster is obtained. The function of $|L_4^a\rangle$, $|L_4^b\rangle$, and $|E_8\rangle$ are already discussed in the main text. The extra qubits in the fourth step are applied for the performing the function of $U_H$ and $R_z$.

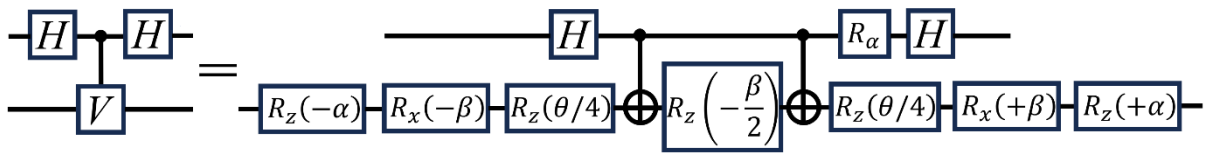

**Fig. S5** A decomposition of the control rotation gate (in Pauli-X basis).

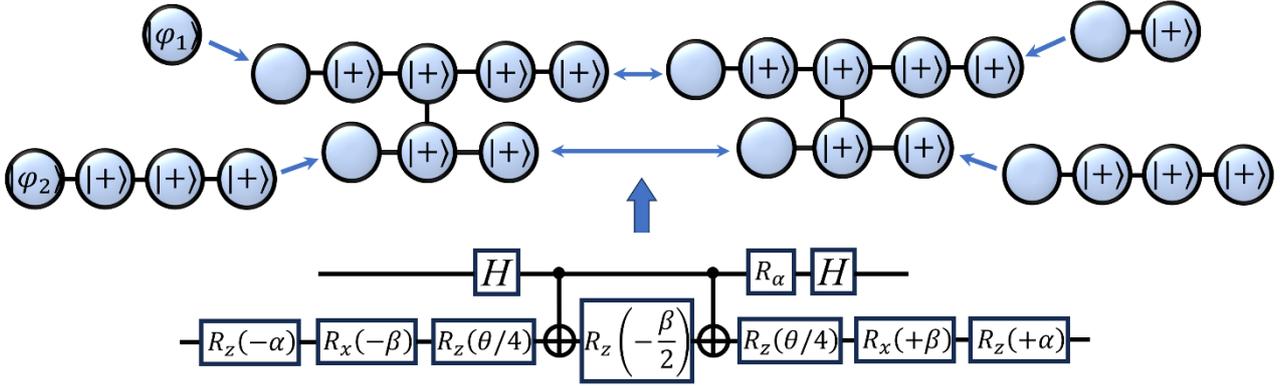

**Fig. S6** The method for obtaining a $V_{ij}$-cluster.

Notice that the state obtained by Fig. S6 is the state shown in Fig. 1(c) of the main text, and the condition for the measurement angles are given by the decomposition circuit in Fig. S5. Like Fig. S4, a graph of the $V_{ij}$-cluster with measurement angles marked out can be given by Fig. S7. The angles satisfy that $a_{51} = -a'_{53}$, $a_{52} = -a'_{52}$, $a_{53} = a'_{51}$, $s = a_{52}/2$, $\varphi = -a_{52}/2$, and the form of the measurement basis is also $\langle 0|R_z(x)$. The structure of a general cluster state for performing the QCNN is given by Fig. 1(f) of the main text. Combining Fig. S4 and S7 and removing the redundant single qubit rotation gates (such as those directly connected to each other, which can be replaced by a single rotation with modified angles), an optimized cluster state corresponding to the state indicated by Fig. 1(f) can be given by Fig. S8.

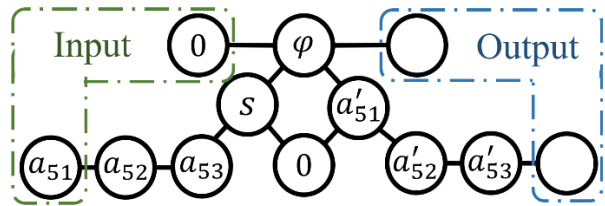

**Fig. S7** An illustration of $V_{ij}$-cluster with all measurement angles marked out.

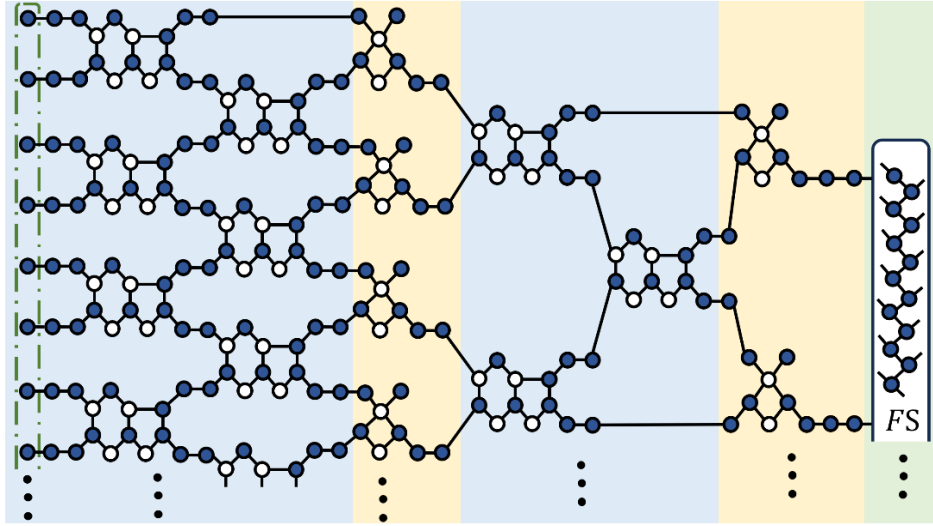

**Fig. S8** An optimized cluster state for performing the function of QCNN. Each small circle represents a qubit in the cluster state. The qubits in the dashed green box represent the input qubits. The blue area marks the qubits for quantum convolutions, which are the convolutional clusters in the main text. The yellow area marks the qubits for pooling, which are the pooling clusters in the main text. The green area marks the qubits for performing the function of a fully connected layer, which is the FC cluster state in the main text and only conceptually displayed. In specific applications, it is a cluster state that performs unitary operations. All the qubits measured by trainable projectors $\langle 0|R_z(x)$ are colored dark blue. The ones measured by fixed projector $\langle 0|$ are colored white.

**S3. The square-lattice for identifying the Haldane ground state.**

As described in the third section of the main text, the square lattice state for identifying the Haldane ground state is graphically shown in Fig. S9(a). The main part of the state is a 2-by-5 square lattices state, and the three qubits $|d_1\rangle$, $|d_2\rangle$, and $|d_3\rangle$ are used for encoding input data. The QCNN proposed by Ref. [3] in such a case can be described by the circuit in Fig. S9(b). $U_a$ and $U_b$ are two unitaries. Control-$V_a$ and control-$V_b$ are two control rotation gates. The exact ansatz of the QCNN circuit in such a case is also given by Ref. [3], which is shown in Fig. S9(c). The three-qubit gate in Fig. S9(c) is a Toffoli gate. To show the relation between the original QCNN circuits and the square lattice state, we use the optimized and strict MBQCNN scheme (shown by Fig. S8) to give the MBQCNN version of the circuit in Fig. S9(b), as shown in Fig. S9(d). Each circle also represents a qubit. The ones measured by $\langle 0|R_z(x)$ are colored dark blue, and the ones measured by $\langle 0|$ are colored white. For the ease of comparison, a version of the state in Fig. S9(a) under the same presentation style is given by Fig. S9(e). From Fig. S9(d) and S9(e), one can conclude that the strict cluster state, which is completely equivalent to the circuit in Fig. S9(b), and the square lattice state share several similarities. They are composed of connected qubits which are measured by trainable projectors, and loop-like structures occur for both of them. Such loop-structures correspond to entangled gates, such as two-qubit or even multi-qubit unitaries, and can guarantee the capability of learning complex data fundamentally. Therefore, they both perform well in the task of identifying the Haldane ground state, as verified by our numerical results. The major differences of the states in Fig. S9(d) and S9(e) are the number of the loop-structures as well as the qubits that are measured by fixed projector $\langle 0|$. In this case, such differences do not cause affections on the results significantly.

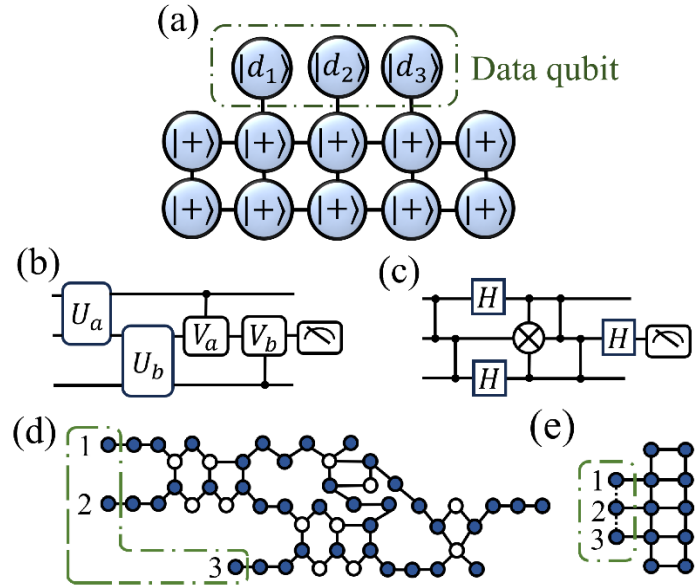

**Fig. S9** The cluster state and the circuits for identifying the Haldane ground state. (a) The square lattice state applied in the first numerical example of the main text. (b) The QCNN circuit in such a case. $U_a$ and $U_b$ are two unitaries. Control-$V_a$ and control-$V_b$ are two control rotation gates. (c) The exact ansatz of the QCNN circuit in such a case. The three-qubit gate in the circuit is a Toffoli gate. (d) The MBQCNN version of the circuit in (b) based on Fig. S8. Qubit 1, 2, and 3 in the dashed green box are the input qubits. (e) A version of the state in (a) under the same presentation style of (d). Qubit 1, 2, and 3 in the dashed green box are also the input qubits.

**S4. The square-lattice for classifying the iris dataset.**

As described in the third section of the main text, the square lattice for classifying the iris dataset is graphically shown in Fig. S10(a). The main part of the state is a 2-by-4 square lattices state with two extra qubits. Each extra qubit is connected to one of the qubits in the cluster state boundaries involving two qubits. Besides, the diagonals of the 2-by-2 square in the other side is also connected for increasing the learning ability of the scheme. The three qubits $|d_1\rangle$, $|d_2\rangle$, $|d_3\rangle$, and $|d_4\rangle$ are used for encoding input data. The CNN and QCNN circuit as the reference for checking the classification performance based on the state in Fig. S10(a) are shown in Fig. S10(b) and S10(c). In the CNN circuit shown in Fig. S10(b), one input data sample (a 16-by-1 vector) is firstly transformed by five convolution kernels, shaped into 14-by-1 vectors. Then, ReLU and average pooling is applied, and a 5-by-1 vector is obtained. Finally, the 5-by-1 vector is fed into a two-layer fully-connected network, and the output can be given. In the QCNN circuit shown in Fig. S10(c), three two-qubit unitaries ($U_1$, $U_2$, and $U_3$) and two control rotations (control-$V_1$ and control-$V_2$) are employed. The details of the networks in Fig. S10(b) and S10(c) are presented in the Appendix A and B of the main text. The trainable parameter numbers of them are the same with that of the MBQCNN based on Fig. S10(a).

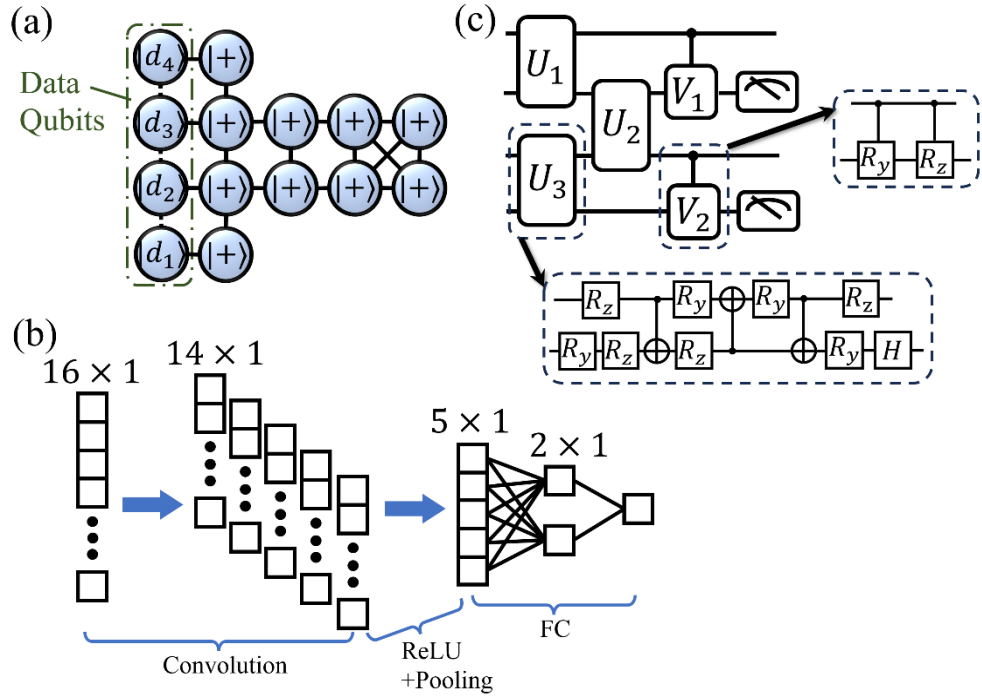

**Fig. S10** The square-lattice for classifying the iris dataset as well as the graphic illustrations of the reference networks. (a) The square lattice state applied in the second numerical example of the main text. (b) The data flow graph of the CNN applied in the second numerical example. The details of the setup are given in Appendix A. (c) The circuit of the QCNN applied in the second numerical example. The details of the setup are given in the Appendix B.